\documentclass[twocolumn,showpacs,preprintnumbers,pre]{revtex4}
\usepackage{amsfonts,amssymb,amsmath,latexsym,epsfig} 
\begin{document}  

\title{Intermittency in two dimensions}

\author{Roberto Artuso } \email{roberto.artuso@uninsubria.it}
\author{Lucia Cavallasca} \email{lucia.cavallasca@uninsubria.it}
\author{Giampaolo Cristadoro } \email{giampaolo.cristadoro@uninsubria.it} 

\affiliation{ Center for Nonlinear and Complex Systems, \\
and Dipartimento di Fisica e Matematica \\ Via Valleggio 11, 22100 Como (Italy); \\
I.N.F.N., Sezione di Milano, Via Celoria 16, 20133 Milano (Italy)}

\date{\today}

\begin{abstract}
We introduce a family of area-preserving maps representing a (non-trivial)  two-dimensional extension of the Pomeau-Manneville family in one dimension. We analyze the long-time behavior of recurrence time distributions and correlations, providing analytical and numerical estimates. We  study the transport properties  of a suitable lift and  use a probabilistic argument to derive the full spectrum of transport moments. Finally the dynamical effects of a stochastic perturbation are considered.
\end{abstract}


\maketitle

\section{Introduction}

Hamiltonian systems are generically not fully hyperbolic \cite{MM}: for example the phase space of  typical area-preserving maps reveals the co-existence of chaotic trajectories and islands of regular motion (periodic or quasi-periodic trajectories) \cite {Mei92}. Even when we are concerned with statistical properties of motion on the chaotic component 
we cannot neglect the presence of regular structures. They deeply influence
chaotic motion as, whenever trajectories come close to integrable islands, they stick there for some time and irregular dynamics is thus punctuated by laminar segments where the system `mimics' an integrable one. This intermittent behavior has strong influence on the long-time properties of quantities like correlations decay or recurrence time statistics, that typically present a power-law tail \cite{lit,lit2,lit3,VCG,gmz}. In unbounded systems intermittency influences  transport properties, generating anomalous diffusion processes (see \cite{ACan} and references therein), in contrast to the normal diffusion observed for fully hyperbolic systems \cite{ll}.
    
While much effort is still devoted to fully understand the general picture of mixed phase space,  the situation simplifies if we let the islands of regular motion shrink to zero: even the presence of a single marginally stable fixed point can produce intermittent-like behavior \cite{ArtPra98, LivMar05,FrGu}.  In one dimension  this corresponds to the  Pomeau-Manneville  maps on the unit interval \cite{PomMan80} 
\begin{equation}
\label{PMmap}
x_{n+1}\,=\,\left. x_n+ x_n^{z}\right|_{mod\,\,1} \qquad z> 1
\end{equation}
which represent one of the few examples of non-fully hyperbolic systems for which analytic results can be obtained with a variety of different techniques (see for instance \cite{Wang,Young,LSV,Zwei,sandro,book,farey,aiz}).
Such maps present a polynomial decay of correlations and recurrence time statistics with exponents that depend on the intermittency parameter $\gamma$ \cite{Young,LSV,Gou}. Moreover a  proper lift on the real line can generate anomalous diffusion and the set of transport moments  typically shows a two-scales structure \cite{theoAN,ark,vulp,ArtCri03}.
    
The situation is much less satisfactory in more than one dimension. Rigorous results were derived for specific 
cases, where it was possible to give precise bounds on the rate of mixing \cite {LivMar05, PolYur01}.  Here we study a family of area-preserving maps with a neutral fixed point. This family depends on a parameter that governs stability properties of the fixed point, in analogy with the Pomeau-Manneville maps. 

We introduce the two dimensional family of area-preserving maps in section II, where we also discuss the dynamical features we will look at. In section III we will consider the unstable manifold of the neutral fixed point, and provide simple estimates that will be pivotal in predicting the decay of survival probabilities. Section IV contains extensive investigations on survival probability and correlation functions decay. By lifting the map on an unbounded phase space we then analyze transport properties in section V. The role of a stochastic perturbation is then studied in section VI, while we present our conclusions in section VII.

\section{The model}

We define the following one-parameter family of maps $T_{\gamma}(x,y): \mathbb{T}^2 \to \mathbb{T}^2$, where  $\mathbb{T}^2=[-\pi,\pi)^2$  (with torus topology):

\begin{equation}\label{themodel}
T_{\gamma}(x,y)=
\left\{ \begin{array}{cc}
x+f_{\gamma}(x)+y & \quad \textrm{on} \, \, \mathbb{T}\\  
y+f_{\gamma}(x)\quad\,\,\,& \quad \textrm{on}\, \, \mathbb{T}
\end{array}\right.
\end{equation}
with

\begin{equation}\label{effe}
f_{\gamma}(x)=\pi \, \textrm{sign}(x)\left|\frac{x}{\pi}\right|^{\gamma}  \qquad \gamma>1.
\end{equation}
\vspace{1cm}

The  map $T_{\gamma}$ is area-preserving for every choice of the impulsive force $f(x)$. The Jacobian matrix is
\begin{equation}\label{Jacobian}
{\bf{J}}_{\gamma}(x,y) = 
\left( \begin{array}{cc} 
1+f'_{\gamma}(x) & 1  \\ 
f'_{\gamma}(x) & 1  
\end{array} \right)
\end{equation}
so we have $\textrm{det}\, {\bf{J}}_{\gamma}(x,y)=1$  and  $\textrm{Tr}\, {\bf{J}}_{\gamma}(x,y)=2+f'_{\gamma}(x) > 2$ for $ x \in  [-\pi,\pi)/\{0\} $: the map is everywhere hyperbolic except on the line $x=0$  and thus the fixed point at the origin is marginally stable (parabolic) for every   $\gamma$. The parameter $\gamma$  changes the `stickiness' of the origin, in analogy with the intermittency exponent $z$ appearing in the Pomeau-Manneville maps of Eq. (\ref{PMmap}).
We point out that the choice   $f(x)=x-\sin(x)$ \cite{ArtPra98, LivMar05} gives rise to a marginal fixed point with the same stability properties of that for  $\gamma=3$ in  Eq. (\ref{themodel}). We note explicitly that  for our model $f \in C^k$ with $k=[\gamma-1] $ (where $[\cdot]$ denotes the integer part) unless $\gamma$ is an odd
integer, in which case $f \in C^{\infty}$ \cite{smnote}.

\subsection{Dynamical indicators}

In order to get information about the  dynamical properties of the systems it is often useful to employ time statistics \cite{M}. We choose a set $\Omega$ including the parabolic fixed point and then define a partition of $\Omega$ in disjoint sets $\Omega_n$, each representing the set of points that leave $\Omega$ in {\em exactly} $n$ iterations.
The survival probability $p_{\Omega}(n)$ is the fraction of initial conditions in $\Omega$ that are still in $\Omega$ after $n$ iterations. The behavior of $p_{\Omega}(n)$ generally depends on the choice of the measure $\mu_i$ with which we distribute initial conditions over $\Omega$, which may  be quite different from the invariant measure. In the present case the invariant measure is the Lebesgue one $\mu$, which also represents the most natural way to spread initial conditions over $\Omega$, so $\mu_i=\mu$ and
\begin{equation}
p_{\Omega}(n)\,=\,\frac{1}{\mu(\Omega)}\sum_{k>n}\, \mu(\Omega_n).
\label{pmu}
\end{equation}
We may also define the waiting time distribution (or residence time statistics) over $\Omega$, $\psi_{\Omega}(n)$, as the probability that once a trajectory enters the set $\Omega$ it stays there {\em exactly} $n$ time steps.  $\psi_{\Omega}(n)$ is computed by running a long trajectory and recording residence times in $\Omega$: $\psi_{\Omega}(n)$ is just the probability distribution of such residence times. In our case
\begin{equation}
\label{psiw}
\psi_{\Omega}(n)\,=\frac{1}{\mu(\Omega)}\left(\mu(\Omega_n)-\mu(\Omega_{n+1})\right)
\end{equation}
where ergodicity and the property that the map preserves Lebesgue measure have been used. We point out that in general, while $p_{\Omega}(n)$ depends upon an -arbitrary- choice of the distribution of initial conditions, $\psi_{\Omega}(n)$ doesn't.

From Eq. (\ref{pmu}, \ref{psiw}) we see that the asymptotics of these quantities are tightly related \cite{M}: in particular if the measure of the sets $\Omega_n$ decays according to a power law 
\begin{equation}
\label{pl-set}
\mu(\Omega_n) \sim n^{-\alpha-1}
\end{equation}
we get
\begin{equation}
\label{pl-surv}
p_{\Omega}(n)\,\sim\, n^{-\alpha}
\end{equation}
and
\begin{equation}
\label{pl-stay}
\psi_{\Omega}(n)\,\sim\, n^{-\alpha-2}.
\end{equation}
Besides their intrinsic interest, these relations bear remarkable links with the problem of establishing the mixing rates for the system \cite{CL,lit2,lit3,DA}. We briefly recall it with a simple argument: suppose we consider an observable $A$ that remains fully correlated for portions of trajectories within $\Omega$, and otherwise completely decorrelated (due to randomness of motion outside of $\Omega$); then we may easily show that \cite{DA}
\begin{eqnarray}
\label{wait-corr}
C_{AA}(n)&=&\langle A(n)A(0)\rangle-\langle A \rangle^2\\ \nonumber 
&\sim& \left(\langle A^{2}\rangle-\langle A \rangle^2\right)\sum_{m=n}^{\infty}\, \sum_{k=m}^\infty\, \psi_{\Omega}(k);
\end{eqnarray}
so that the exponents of power-law decay of survival probability and correlations should coincide
\begin{equation}
\label{pl-corr}
C_{AA}(n) \,\sim\, n^{-\alpha}.
\end{equation}
We observe that the validity of Eq. (\ref{wait-corr}) has been carefully numerically scrutinized for chaotic billiards \cite{acgJSP,DA}, and even validated in rigorous estimates of polynomial mixing speed for $1d$ intermittent systems \cite{Young} (see also \cite{StIs}), but indications of its possible failure have also been suggested \cite{G-T}.

It is interesting to remark that showing that power law decays of $p_{\Omega}(n)$ and $\psi_{\Omega}(n)$ differ by two (Eq. (\ref{pl-surv}, \ref{pl-stay})) employs the fact that Lebesgue measure is the invariant one for the system. Actually for the map of Eq. (\ref{PMmap}), where the invariant measure is not uniform, they differ by one if we choose the initial conditions uniformly distributed with Lebesgue measure (while the exponent of the waiting time distribution coincides with the one ruling the length of the segments $\Omega_n$ \cite{PomMan80}) . A relationship coinciding with Eq. (\ref{pl-surv}, \ref{pl-stay}) holds instead for another intermittent map, introduced by Pikovsky \cite{ark} (some features of this map are also described in \cite{ACJP}):
\begin{equation}
\label{zArk}
x_{n+1}\,=\,\tilde{f}_z(x_n),
\end{equation}
where $\tilde{f}_z$ is an (odd) circle map, again dependent on an intermittency parameter $z$, implicitly defined on the torus $\mathbb{T}=[-1,1)$ by
\begin{equation}
x\,=\,
\left\{
\begin{array}{ll}
\frac{1}{2\gamma}\left(1+ \tilde{f}_z(x)\right)^{z} \qquad & 0 < x < 1/(2z) \\
\tilde{f}_z(x) + \frac{1}{2z} \left( 1- \tilde{f}_z(x) \right)^{z} \qquad & 1/(2z) < x < 1
\end{array}
\right.
\label{arkmap}
\end{equation}
A key feature that the map of Eq. (\ref{zArk}) shares with our model is that the invariant distribution is smooth, coinciding with the Lebesgue measure, as it can be checked by inspection of the form of Perron-Frobenius operator. 

For the map under consideration it is possible to get an estimate of the exponent $\alpha$  by studying the invariant manifolds of the parabolic fixed point. 

\section{Invariant manifolds}

\begin{figure}[!htb]
\begin{center}
\includegraphics[width=\columnwidth]{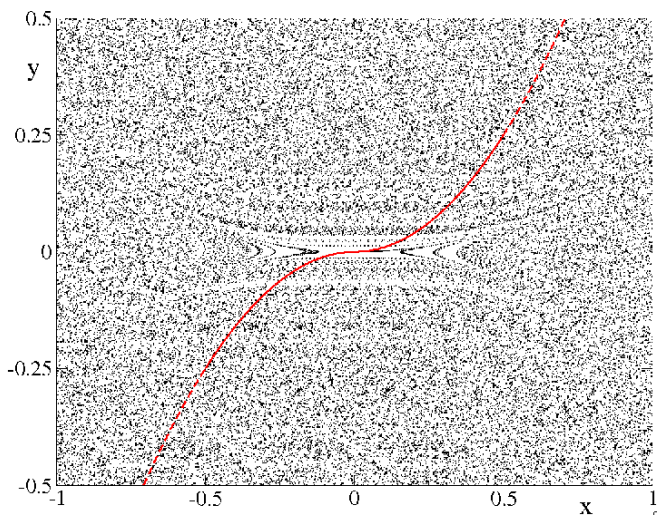}
\end{center}
\caption{\small{(color online) Portion of phase space of the map of Eq. (\ref{themodel}) for  $\gamma=3$  close to the marginal fixed point together with the graph of  its unstable manifold (continuous (red) line)}.}
\label{mappo}
\end{figure}

A typical trajectory  staying for a long time in $\Omega$ (again we take as $\Omega$ a region including the parabolic fixed point) enters $\Omega$ close to the stable manifold, escaping along the unstable one (Fig. (\ref{mappo})).  In particular we are going to discuss how trajectories escape by following the unstable manifold of the marginal fixed point. For the odd symmetry of $f_{\gamma}(x)$, we can restrict the analysis to the first quadrant.

Let's call $(x,y(x))$ the graph of the unstable manifold in the neighborhood of the origin and suppose that, very close to the indifferent fixed point, $y(x)\simeq x^{\sigma}$.
With the choice of Eq. (\ref{effe}) we have that $f_{\gamma}(x) \simeq a \; x^{\gamma}$ and  the Jacobian matrix  (Eq. (\ref{Jacobian})) of the map is:
\begin{equation}
\textbf{J}_\gamma(x,y)=\left(
\begin{array}{ccc}
1+b\;x^{\gamma-1}&&1\\
b\;x^{\gamma-1}&&1\\
\end{array}
\right)
\end{equation}
whose eigenvalues are written to leading order as:
$\lambda_{\pm}=1\pm \beta \; x^{(\gamma-1)/2}$.\\
The vector  $(1,y'(x))$, \emph{i.e.}  $(1, \eta\;x^{\sigma-1})$, tangent to the unstable manifold satisfies 
\begin{equation}
\textbf{J}_\gamma(x,y)
\left(
\begin{array}{c}
1\\
\eta\;x^{\sigma-1}
\end{array}
\right)\simeq(1+\beta \; x^{(\gamma-1)/2})
\left(
\begin{array}{c}
1\\
\eta\;x^{\sigma-1}
\end{array}
\right)
\end{equation}
i.e.
\begin{equation}
\left\{
\begin{array}{rcl}
1+b\;x^{\gamma-1}+\eta\;x^{\sigma-1}&\simeq& 1+\beta\;x^{(\gamma-1)/2}\\
b\;x^{\gamma-1}+\eta\;x^{\sigma-1}&\simeq&\eta\;x^{\sigma-1}+\eta\beta\;x^{(\gamma-1)/2+(\sigma-1)}\\
\end{array}
\right.
\end{equation}
and from these equations, remembering that  $x<<1$ and $\gamma>1$, we obtain
\begin{equation}\label{manifold}
\sigma=\frac{\gamma+1}{2}.
\end{equation}
We note explicitely  that for $\gamma=3$, this result is in agreement with the case $f=x-\sin(x)$ derived in  \cite{LivMar05}.
We can now study the dynamics restricted to the unstable manifold. Let's call $\ell$ the arclength coordinate along the manifold; for small $x$ we get:
\begin{equation}\label{eqmanifold}
\ell(x)\,=\int^{x}dx\sqrt{1+(dy(x)/dx)^{2}}\simeq x
\end{equation}
Denote by $\ell_{n}$ the coordinate $\ell$ at a point $\left( x_n,y(x_n) \right)$ and by $\ell_{n+1}$ the coordinate along the manifold of $T_{\gamma}\left(x_n,y(x_n)\right)$
\begin{equation}
\ell_{n+1}=\ell_{n}+h(\ell_{n}).
\end{equation}
By using Eq. (\ref{eqmanifold}) and (\ref{themodel}) we get
\begin{equation}
h(\ell)\simeq\frac{d\ell}{dt}(\ell)=\frac{d\ell}{dx}\frac{dx}{dt}(\ell)\simeq  (y(\ell)+x^{\gamma}(\ell))=\ell^{\sigma}+\ell^{\gamma}
\end{equation}
From Eq. (\ref{manifold}) and by remembering that $\gamma > 1$ we obtain, via a continuous time approximation \cite{PS,hirsch},
\begin{equation}
h(\ell)\simeq \ell^{\sigma}=\frac{d\ell}{dt}.
\end{equation} 
If we fix the boundary of $\Omega$ at a scale $L$ we can then evaluate the time needed to reach the boundary as a function of the arclength $\ell$ along the manifold, by employing the standard argument of \cite{PS,hirsch}:
\begin{equation}
\label{cT}
T(\ell)\,=\,\frac{2}{\gamma-1}\left( \ell^{-\frac{\gamma-1}{2}} -L^{-\frac{\gamma-1}{2}}\right);
\end{equation}
that implies the following scaling for the inverse function
\begin{equation}
\label{invcT}
\ell(T)\,\sim\,T^{-\frac{2}{\gamma-1}}.
\end{equation}
We now arrive to the crucial point: we estimate $p_{\Omega}(n)$ as the area of rectangle having one vertex at the origin (the parabolic fixed point), and another at a point on the unstable manifold $(\overline{x},\overline{y})$ that exits $\Omega$ in $n$ steps (see Fig. ({\ref{part})):
\begin{equation}
p_{\Omega}(n)\,\simeq\,\overline{x}\cdot \overline{y}\,\simeq \, \ell(n)^{\sigma+1}\simeq (n^{-\frac{2}{\gamma-1}})^{\sigma+1}\,=\,n^{-\frac{\gamma+3}{\gamma-1}}
\label{xy}
\end{equation}
\begin{figure}[!htb]
\begin{center}
\includegraphics[width=\columnwidth]{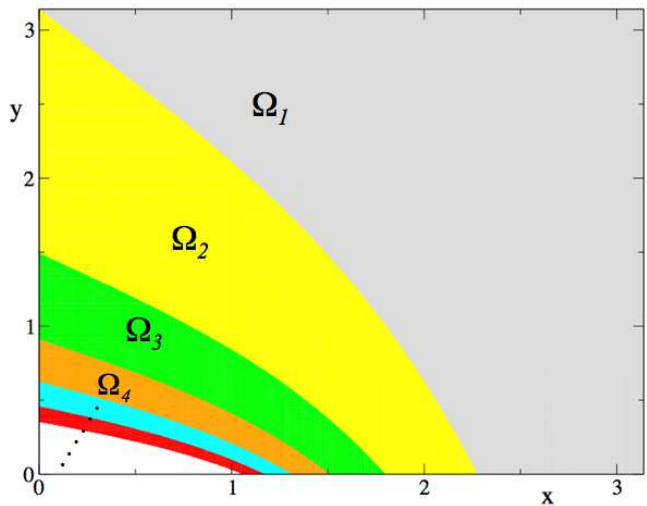}
\end{center}
\caption{\small{(color online) A few $\Omega_n$ (once we set $\Omega$ as the first quadrant $x \geq 0$, $y \geq 0$)}.}
\label{part}
\end{figure}
Thus we have an estimate of power law decay of the survival probability as a function
of the intermittency parameter $\gamma$ as
\begin{equation}
\label{pn}
p_{\Omega}(n)\,  \sim\, n^{-\frac{\gamma+3}{\gamma-1}}
\end{equation}
and for the waiting time distribution as well (Eq. (\ref{pl-stay}))
\begin{equation}
\label{psin}
\psi_{\Omega}(n)\, \sim \, n^{-\frac{3\gamma+1}{\gamma-1}}.
\end{equation}
In view of the argument we earlier mentioned (see Eq. (\ref{pl-corr})), the estimate of Eq. (\ref{pn}) suggests the same decay law for (auto)correlation functions
\begin{equation}
\label{pcorr}
C_{AA}(n)\,\sim\, n^{-\frac{\gamma+3}{\gamma-1}}.
\end{equation}
Next section will present several numerical simulations concerning these quantities.

\section{Asymptotic decays}
We start by considering the survival probability:
Fig. (\ref{survo}) shows two examples of numerically computed $p_{\Omega}(n)$. 
The numerical data exhibit an excellent agreement with analytic estimates over a wide range of intermittency parameters, as shown in Fig. (\ref{exponento}), which also provides clear indications of the validity of our estimate for the asymptotic decay of the waiting time distribution.

\begin{figure}[!htb]
\begin{center}
\includegraphics[width=\columnwidth]{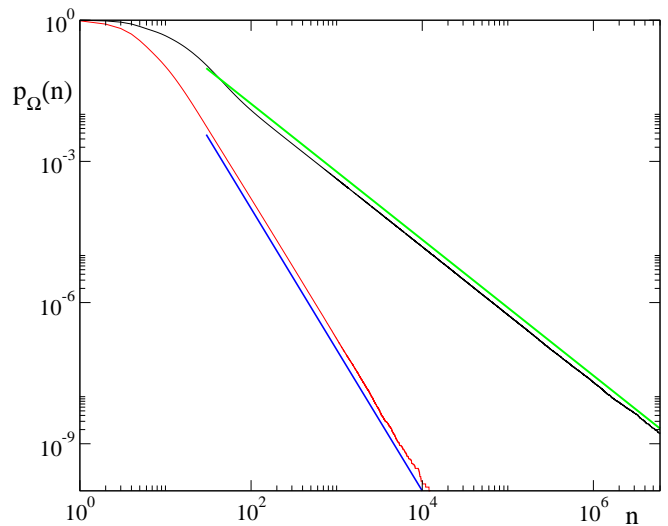}
\end{center}
\caption{\small{(color online) Survival probabilities for $\gamma=3$ (lower curve) and $\gamma=10$ (upper curve) together with the  power law decays  predicted by Eq. (\ref{pn}). We used $10^{12}$ initial conditions and set $\Omega=\left[ -\frac{1}{2},\,\frac{1}{2}\right]^{2}$ }. }
\label{survo}
\end{figure}

\begin{figure}[!htb]
\begin{center}
\includegraphics[width=\columnwidth]{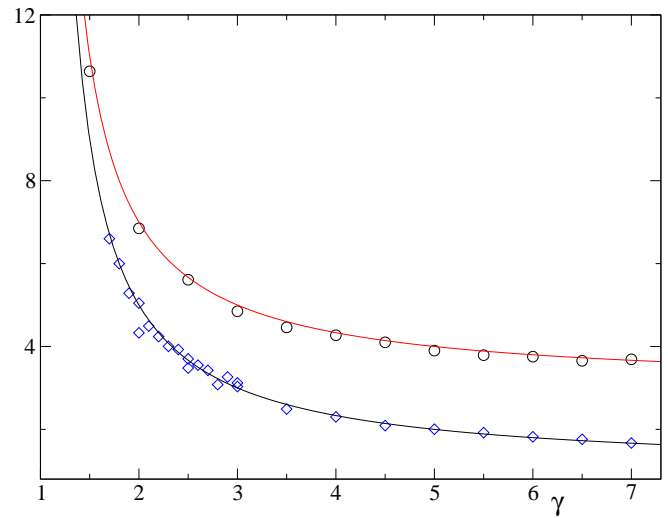}
\end{center}
\caption{\small{(color online) Exponents of power-law decay for survival probability $p_{\Omega}(n)$ and waiting time distribution $\psi_{\Omega}(n)$: lines refer to analytic estimates of Eq. (\ref{pn})-upper, (\ref{psin})-lower: open circles come from numerical simulations of the waiting time distribution, open diamonds from survival probability simulations}.}
\label{exponento}
\end{figure}

We already mentioned that various arguments support the expectation that correlation functions should decay as the survival probability (Eq. (\ref{pcorr})), so we proceed to scrutinize this prediction by running extensive direct numerical simulations on autocorrelation functions; as we are dealing with an ergodic (and mixing \cite{LivMar05}) system,  autocorrelation functions can be evaluated in terms of phase space averages (instead of temporal averages):
\begin{equation}
C_{AA}(n)=\int_{\mathcal{M}}d\mu(z)A(T_{\gamma}^{n}z)A(z)-\left(\int_{\mathcal{M}}d\mu(z)A(z)\right)^{2}
\label{corr}
\end{equation}
where $A$ is a smooth function on the phase space $\mathcal{M}$ and $\mu$ is the invariant Lebesgue measure.  
From a numerical point of view it is known that often Monte-Carlo evaluation of Eq. (\ref{corr}) cannot be pushed too far, as the statistical error is of order $1/\sqrt{N}$ in the number of initial conditions: generally we also expect an (exponential) transient in the decay \cite{ArtPra98,DA,per-ep}: transient time $\overline{t}$ might depend on both $\gamma$ and the choice of phase space function $A$ \cite{VCG}. We also remark that smoothness of the function $A$ plays a fundamental role: as a matter of fact we may obtain arbitrarily slow correlation decay even for Anosov maps by using integrable non-smooth functions \cite{CC}, or, from a mathematical point of view, we may have that the degree of smoothness determines the essential spectral radius of the Perron-Frobenius operator \cite{CI}.

We performed the explicit calculation of the autocorrelation function for different values of the intermittency parameter $\gamma$ and for different observables. We obtained the best results (i.e. cleanest curves and shortest time $\overline{t}$) for large values of $\gamma$ and by using $A(x,y)=e^{-y^{2}}$. The choice of the smooth function to use is quite arbitrary; we looked for a function not vanishing in correspondence of the marginal fixed point (as suggested for example in \cite{VCG}) and by the special choice of a gaussian depending on a single variable we could save computational time (see also \cite{aiz}).
In figures (\ref{corr3}) and (\ref{corr10}) we present results for $\gamma=3$ and $\gamma=10$: 

\begin{figure}[!ht]
\begin{center}
\includegraphics[width=\columnwidth]{5.eps}
\end{center}
\caption{\small{(color online) Autocorrelation function for the observable $e^{-y^{2}}$ and $\gamma=3$. We used $2\cdot10^{10}$ initial conditions (uniformly distributed in the torus cell). The predicted decay is shown by the (red) straight line, which has a slope $-3$}.}
\label{corr3}
\end{figure}
\begin{figure}[!ht]
\begin{center}
\includegraphics[width=\columnwidth]{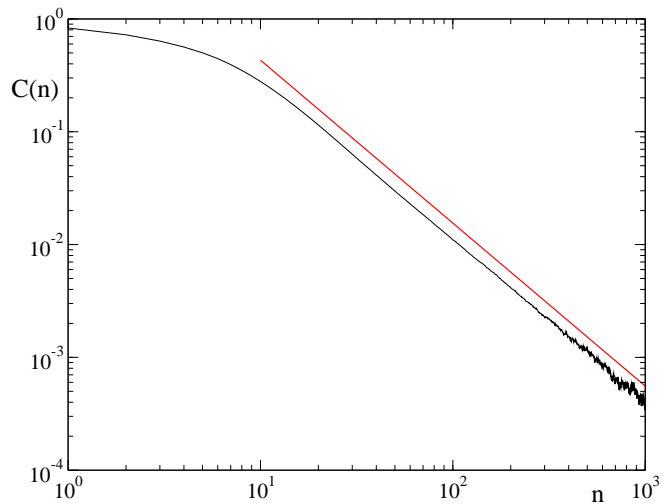}
\end{center}
\caption{\small{(color online) Autocorrelation function for the observable $e^{-y^{2}}$ and $\gamma=10$. We
used $10^{9}$ initial conditions (uniformly distributed in the torus cell). The predicted decay is shown by the (red) straight line, which has a slope $-13/9$}.}
\label{corr10}
\end{figure}
The case reported in Fig. (\ref{corr3}) is important because for a similar $2d$ mapping (having the same intermittent exponent $\gamma=3$) it was proved in \cite{LivMar05} that the decay is faster than $n^{-2}$, and a class of cross-correlation was constructed indicating that the bound is optimal: while survival probability data for the same exponent indicate clearly that the decay we predict ($n^{-3}$) is numerically well reproduced, data for correlations are less conclusive (see Fig. (\ref{corr3})). In general, numerical data are more difficult to interpret for low values of $\gamma$, and numerical fits tend to lie below the predicted exponents (see Fig. (\ref{exponent-corr})), while for larger values of $\gamma$ the accordance with our estimates is much better. Moreover, the agreement improves by increasing the number of initial conditions, so that we expect the discrepancy to be essentially due to numerical limitations (Fig. (\ref{exponent-corr})).
\begin{figure}[!ht]
\begin{center}
\includegraphics[width=\columnwidth]{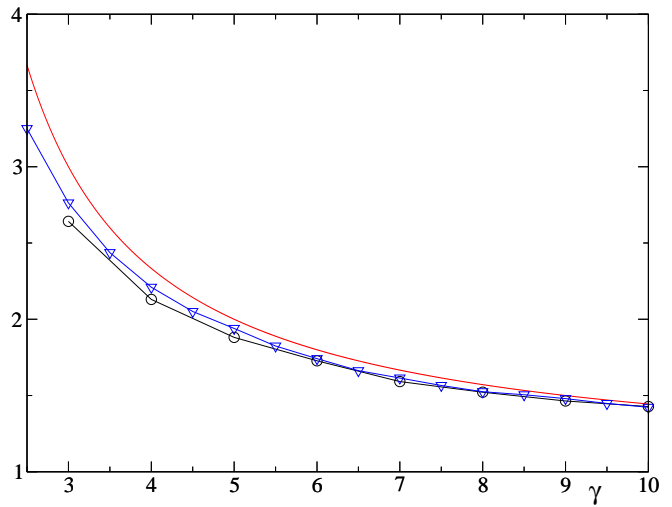}
\end{center}
\caption{\small{(color online) Numerical values of power-law decay exponents for the autocorrelation function for the observable $e^{-y^{2}}$ ((black) circles and (blue) triangles) together with the analytical estimates (full (red) line). Circles were obtained by using $2\cdot10^{9}$ initial conditions while the triangles were obtained for $\gamma=2.5$ by using $5\cdot10^{10}$ initial conditions, in the range $3 \leq \gamma \leq 6$ by using $2\cdot 10^{10}$ initial conditions, and in the range $6.5 \leq \gamma \leq 10$ by using $10^{10}$ initial conditions}.}
\label{exponent-corr}
\end{figure}
Further indications of the similarity between correlations and survival probability decays will be provided in section VI, when considering the role of stochastic perturbations.

\section{Transport properties}

In order to study transport properties, we have to abandon the dynamics restricted to the torus  (Eq. (\ref{themodel}))  by lifting the map in an appropriate way.
 
For the sake of clarity we introduce a third dimension (say $z$) to describe the motion through elementary cells. We then assign a \textit{jumping number} $+1$ to the points belonging to the first quadrant, $-1$ to the points belonging to the third quadrant and $0$ to all the other points. This means that a laminar phase of length $n$ will correspond to a jump in the positive direction of $n$ elementary cells (Fig.\ref{jumping}).
\begin{figure}[ht!]
\includegraphics[width=\columnwidth]{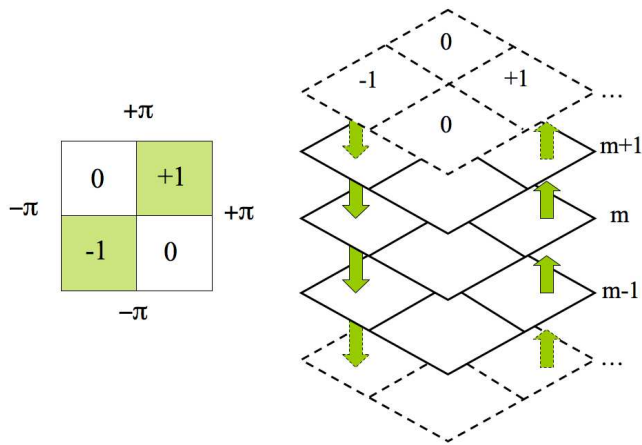} 
\caption{\small{Jumping numbers and lifted map}.}
\label{jumping}  
\end{figure}

Formally, the lift is given by the following formula:
\begin{eqnarray}\label{themodeltra}
\overline{T}_{\gamma}(x, y, m)=
\left\{ \begin{array}{cc}
(T_{\gamma}(x, y), m)& \quad \textrm{for}\, xy < 0 \\
(T_{\gamma}(x,y), m+ \textrm{sign}(x)) & \quad \textrm{for} \, xy\ge0
\end{array}\right.
\end{eqnarray}
where $m$ is an integer variable.

Considering successive entrance in the laminar regions as uncorrelated we can approximate the diffusion process by a Continuous Time Random Walk (CTRW) \cite{sm, z-k}, with  the probability distribution of the laminar phases given by the waiting time distribution  $\psi_{\Omega}(n)$.
\\In particular, by making use of the CTRW approach it is possible to characterize the transport properties of the process  in terms of the set of moments of the diffusing variable \cite{vulp-anders}:
\begin{equation}
\langle \left|z(n)-z(0)\right|^q \rangle \simeq n^{\nu(q)}
\end{equation}
that is expected to present a sort of phase transition \cite{vulp}. 

\subsection{Continuous Time Random Walk approach}
For completeness, we briefly recall the standard theory of Continuous Time Random Walks \cite{sm, z-k, vulp-anders, sokolov, k-b-s}.
Generally speaking, a CTRW is a stochastic model in which steps of a simple random walk take place at times $t_{i}$, following some waiting time distribution. Mathematically, it is asserted that a CTRW is a (non-Markovian) process subordinated to a random walk under the operational time defined by the process ${t_{i}}$ \cite{sokolov}.

A CTRW is completely characterized by the quantity $\psi(r,\tau)$, the probability density function to move a distance $r$ during a time interval $\tau$ in a single motion event; the dependence upon $r$ and $\tau$ can be either decoupled (i.e. $\psi(r,\tau)=\chi(r)\wp(\tau)$) or coupled (e.g. $\psi(r,\tau)=\chi(r|\tau)\wp(\tau)$).

The object we are interested in, is the probability density function $P(x,t)$ of being in $x$ at time $t$; indeed it allows us to obtain the full spectrum of transport moments, through the formula
\begin{equation}\label{inverselaplace}
\langle x(t)^{q}\rangle = (i)^{q} L^{-1}\left[ \frac{\partial^{q}}{\partial k^{q}}\tilde{\hat{P}}(k,u)\arrowvert_{k=0} \right]
\end{equation}
where $L^{-1}$ is the inverse Laplace transform and $\tilde{\hat{P}}$ denotes the Fourier-Laplace transform, being $k$ the Fourier variable and $u$ the Laplace variable.
\\Let's introduce $\phi(x,t)$, the probability density function of passing through $(x, t)$, even without stopping at $x$, in a single motion event
\begin{equation}
\phi(x,t)=P(x|t)\int_{t}^{\infty} \!\! d\tau \!\! \int_{|x|}^{\infty} \!\! dr\, \psi(r,\tau).
\end{equation}
$P(x,t)$ is given by the sum of the probabilities of passing  through $(x, t)$, even without stopping at $x$, in one or more motion events 
\begin{equation}
P(x,t)=\phi(x,t)+\int_{-\infty}^{\infty} \!\! dx' \!\! \int_{0}^{t} \!\! d\tau\, \psi(x',\tau)\phi(x-x',t-\tau)+\dots
\end{equation}
By performing the convolution integrals, the Fourier-Laplace transform of this expression assumes the closed form
\begin{equation}
\label{P-F-L}
\tilde{\hat{P}}(k,u)=\frac{\tilde{\hat{\phi}}(k,u)}{1-\tilde{\hat{\psi}}(k,u)}
\end{equation}

A special realization of CTRW is the so called \textit{velocity model} \cite{z-k}: a particle moves at a constant velocity for a given time, then stops and chooses a new direction and a new time of sojourn at random according to given probabilities.
\\Our case belongs to this class, with velocities being $\pm1$, and
\begin{equation}
\chi(r|\tau)=\frac{1}{2}\delta(|r|-\tau) \quad \textrm{and} \quad \wp(\tau)\sim\tau^{-g}
\end{equation}
so that
\begin{equation}
\psi(r,\tau)\sim\frac{1}{2}\delta(|r|-\tau)\tau^{-g} \quad \textrm{and} \quad \phi(x,t)\sim\frac{1}{2}\delta(|x|-t)t^{-g+1}
\end{equation}
where $g=\frac{3\gamma+1}{\gamma-1}$, being $\wp(\tau)$ given by the waiting time distribution $\psi_{\Omega}(n)$ of Eq. (\ref{psin}).

By making use of the Tauberian theorems for the Laplace transform \cite{Feller} and by applying the CTRW formalism \cite{k-b-s} we derive, through Eq. (\ref{inverselaplace}) and  (\ref{P-F-L}) the full spectrum of transport moments.

The obtained spectrum of moments (more precisely, from the previous calculation it is possible to obtain only the \textit{even} moments, and then to infer that a similar law drives also the behavior of the absolute value of \textit{odd} moments) is: 
\begin{equation}
\langle \left|z(n)-z(0)\right|^q \rangle \simeq n^{\nu(q)}
\end{equation}
where the exponent $\nu(q)$ has a piecewise linear behavior
\begin{equation}\label{spectrum}
\nu(q) = \left\{ \begin{array}{ll} 
q/2 & \textrm{if $q<2\alpha$}\\ 
q-\alpha & \textrm{if $q>2\alpha$} 
\end{array} \right. 
\quad \quad \alpha=\frac{\gamma+3}{\gamma-1}
\end{equation}
in agreement with numerical results shown in Fig. (\ref{moment-MIA}).
\begin{figure}[ht!]
\includegraphics[width=\columnwidth]{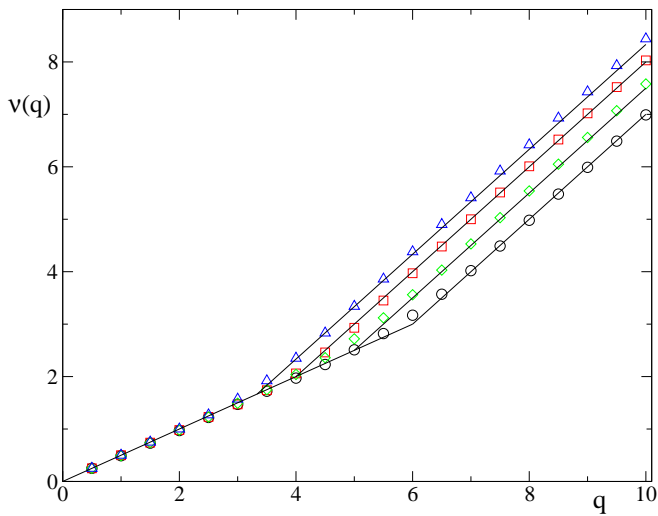}
\caption{\small{(color online) Spectrum of the transport moments for different value of the parameter $\gamma$. Lines correspond to theoretical predictions of Eq. (\ref{spectrum}), symbols correspond to numerical simulations: circles $\gamma=3$,  diamonds $\gamma=11/3$, squares $\gamma=5$, triangles $\gamma=7$}.}
\label{moment-MIA}  
\end{figure}
The transition at $q=2\alpha$ in momenta spectrum of Eq. (\ref{spectrum}) is general in systems manifesting anomalous diffusion \cite{vulp}.

As an outcome, we have that (anomalous) transport properties fully agree with the power laws we deduced for the waiting time distribution (Eq. (\ref{psin})).

\section{Noise effects}

In order to better understand the link between correlations decay and time statistics (and to verify it, if not rigorously prove), we consider the effects of a small stochastic perturbation. The behavior, under the modified dynamics, of the survival probability and of correlations decay may provide further informations about the interconnection between them. At the same time the dynamical effects of a superimposed noise are interesting by themselves (see references \cite{LSV,hirsch, 
mannella, altmann}).
\\The general expectation is that small scale stochasticity blurs the behavior in the vicinity of the parabolic fixed point, enhancing the chaotic character of motion; one then expects a transition to an exponential decay of the survival probability and correlation functions: this intuition is corroborated by numerical experiments, reported in Fig. (\ref{corr-noise}).
\begin{figure}[ht!]
\includegraphics[width=\columnwidth]{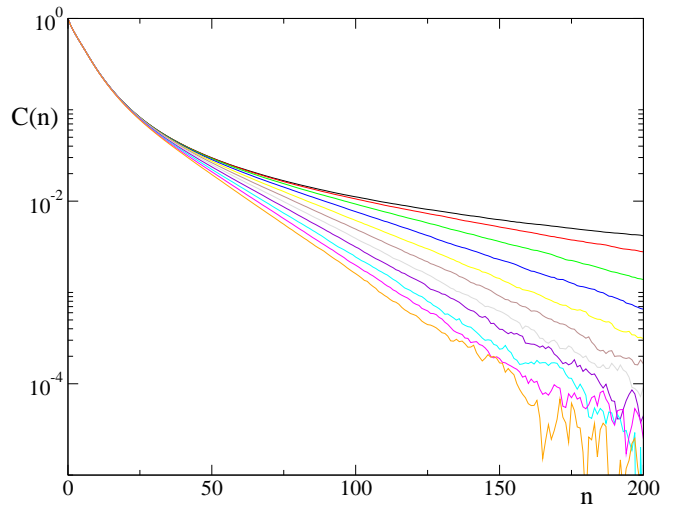} 
\caption{\small{Correlation decay for noisy dynamics, for $\gamma=10$ and various values of
$\epsilon$ (from top to bottom: $\epsilon=0.0$, $\epsilon=0.005$, $\epsilon=0.010$, $\epsilon=0.015$, $\epsilon=0.020$, $\epsilon=0.025$, $\epsilon=0.030$, $\epsilon=0.035$, $\epsilon=0.040$, $\epsilon=0.045$, $\epsilon=0.045$, $\epsilon=0.050$). Each correlation function is computed by considering $3\cdot 10^9$ initial conditions}.}
\label{corr-noise}  
\end{figure}
We perturb the system by introducing a stochastic noise, adding at each iteration of the map a  random vector of the type $\xi=(\xi_1,\xi_2)$ with $\xi_i$   i.i.d in $(-\epsilon ,\epsilon)$.
The effects of the perturbation are expected to be dominant in the region of phase space (that will depend on the noise intensity $\epsilon$ and on the stickiness parameter $\gamma$) where the deterministic step is small compared to noise. Basing on this assumption, we are able to give an analytical estimate of the crossover time $t_c$ (defined as the characteristic time of the asymptotic exponential decay), which is in good agreement with numerical simulations. 

Following \cite{mannella} we divide the phase space into two complementary sections: one small region surrounding the fixed point, where the dynamics is dominated by stochastic diffusion, and its complementary, far from the fixed point, where the dynamics is dominated by the deterministic chaotic motion. 
\\Now the main problem is a proper definition of the boundary of such a partition.
The criterion suggested in \cite{mannella} consists in defining $\langle T_{deterministic}\rangle$ as the mean exit time from the region, say $\Delta$, evaluated with the assumption that the dynamics is only due to the deterministic motion of the unperturbed map; then we define $\langle t_{random}\rangle$ as the mean exit time evaluated as if the dynamics were only due to diffusion. \\The border of the region is determined by the constraint:
\begin{equation}
\langle T_{deterministic}\rangle\simeq\langle t_{random}\rangle.
\label{tT}
\end{equation}

We restrict the analysis to the first quadrant and choose as region $\Delta$ the area defined by the survival probability $p_{\Omega}(k)$, for some time $k$.
In this way $\langle T_{deterministic}^{k}\rangle$ is given by
\begin{equation}
\langle T_{deterministic}^{k}\rangle=\frac{1}{p_{\Omega}(k)}\sum_{n\ge k}^{\infty}\,\mu(\Omega_{n})\cdot (n-k+1).
\label{t-det-1}
\end{equation}
Performing the calculation by substituting the probabilities obtained in the previous sections (Eq. (\ref{pl-set}, \ref{pl-surv})), we get
\begin{equation}
\langle T_{deterministic}^{k}\rangle\simeq k.
\label{t-det-2}
\end{equation}

The calculation of $\langle t_{random}\rangle$ is performed as follows: firstly we approximate  $p_{\Omega}(k)$ as in section III with the rectangular regions of Eq. (\ref{xy}), say $x_{k}\cdot y_{k}$; these rectangular region can be exited along the $x-$direction or along the $y-$direction, independently (thanks to the particular form of our noise), so that
\begin{equation}
\langle t_{random}^{k}\rangle=\min \left(\langle t_{random}^{k,\,x}\rangle,\,\langle t_{random}^{k,\,y}\rangle\right).
\label{t-ran-1}
\end{equation}
From the diffusion equation describing stochastic dynamics (see \cite{mannella,agmon}) we get
\begin{equation}
\langle t_{random}^{k,\,z}\rangle\simeq\frac{z_{k}^2}{\epsilon^{2}},
\label{t-ran-2}
\end{equation}
where $z$ can be either $x$ or $y$. Remembering that close to the origin the motion follows dynamics on the unstable manifold and from Eq. (\ref{invcT}, \ref{xy}) we get
\begin{equation}
\begin{array}{ccccc}
x_{k}&\sim &\ell_{k}&\sim &k^{-\frac{2}{\gamma-1}}\\
y_{k}&\sim &x_{k}^{\sigma}&\sim &k^{-\frac{\gamma+1}{\gamma-1}}.\\
\end{array}
\label{t-ran-3}
\end{equation}
By substituting Eq. (\ref{t-ran-3}) in Eq. (\ref{t-ran-2}) we obtain
\begin{equation}
\langle t_{random}^{k}\rangle\simeq \min \left(\frac{(k^{-\frac{2}{\gamma-1}})^2}{\epsilon^{2}}, \, \frac{(k^{-\frac{\gamma+1}{\gamma-1}})^2}{\epsilon^{2}}\right).
\label{t-ran-4}
\end{equation}
and finally, from Eq. (\ref{tT})
\begin{equation}
k\sim\frac{(k^{-\frac{\gamma+1}{\gamma-1}})^2}{\epsilon^{2}}.
\end{equation}
Writing the characteristic time of the exponential decay as a function of the noise strength $\epsilon$ and the intermittency parameter $\gamma$, we derive an estimate for the crossover time $t_{c}$:
\begin{equation}
t_{c}=k\sim\epsilon^{-\beta(\gamma)}\sim\epsilon^{-\frac{2\gamma-2}{3\gamma+1}}.
\label{cross-time}
\end{equation}

Eventually we compare the behavior of $\beta(\gamma)$ with the numerical results, either for the survival probability and for the correlations decay.

Numerical results are obtained in the following way: for each value of $\gamma$ we consider either the survival probability or correlation functions for several values of $\epsilon$ (see Fig. (\ref{corr-noise}): the exponential decay rates, that we consistently observe, are then fitted according to a power-law in $\epsilon$.

The nice resemblance between the numerical simulations for the correlations decay and the survival probability, together with a good agreement with the analytical result of Eq. (\ref{cross-time}) (see Fig. \ref{surv-corr}) strengthen our belief that the two distributions in the unperturbed case should be driven by the same exponent.

\begin{figure}[ht!]
\includegraphics[width=\columnwidth]{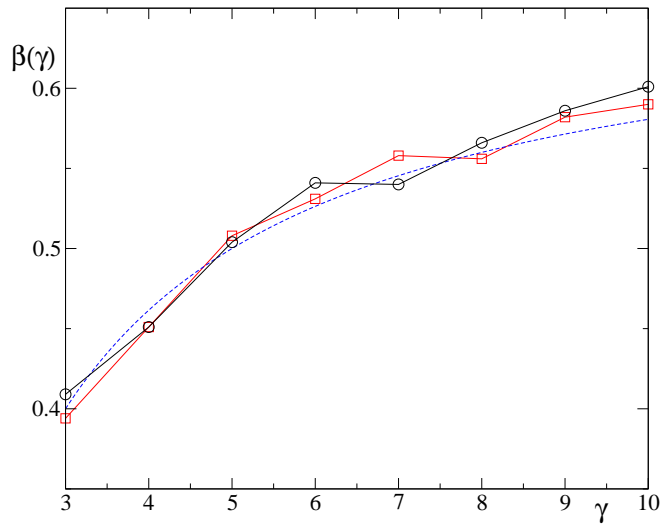} 
\caption{\small{(color online) Comparison between theory and numerical results for the function $\beta(\gamma)$ of Eq. (\ref{cross-time}). Circles: numerical data for the survival probability; squares: numerical data for the correlations; dashed line: theoretical prediction}.}
\label{surv-corr}  
\end{figure}

\section{Conclusions}
We considered a one-parameter family of area-preserving maps which generalizes intermittent behavior in two dimensions (while admitting a smooth invariant measure). This is a paradigmatic example of weak chaos for hamiltonian maps even if the measure of the ``regular" portion of the phase space is zero (only a parabolic fixed point): the parameter $\gamma$ controls sticking of trajectories to the fixed point. By considering the motion along the unstable manifold close to the fixed point we are lead to estimates for power law decays of the survival probability and waiting time distribution. Numerical computations of survival probabilities and residence time statistics show a very close agreement with predicted exponents. Correlation functions are harder to deal with numerically, yet in this case results are close to analytic predictions. Our results are further supported either by considering transport properties for a lift of the map, and by studying dynamical effects induced by a stochastic perturbation.

While we think that a complete -quantitative- understanding of weak chaos in more that one dimension is still in its infancy, our results shed new light on connecting local features (motion along the unstable manifold close to the fixed point) to global dynamical quantities, as mixing speed, transport properties, and response to stochastic perturbations.

This work has been partially supported by MIUR--PRIN 2005 projects {\em 
Transport properties of classical and quantum systems} and {\em Quantum computation with trapped particle arrays, neutral and charged}. 
We thank Raffaella Frigerio and Italo Guarneri for sharing their 
early work on the problem, and Carlangelo Liverani and Sandro Vaienti for useful discussions and informations.

\end{document}